# Non-resonant n = 1 helical core induced by m/n = 2/1 tearing mode in JT-60U

T. Bando[1], S. Inoue[1], K. Shinohara[1,2], A. Isayama[1], T. Wakatsuki[1], M. Yoshida[1], M. Honda[1],

G. Matsunaga[1], M. Takechi[1], N. Oyama[1], and S. Ide[1]

[1]National Institutes for Quantum and Radiological Science and Technology, Naka, Ibaraki 311-0193, Japan

[2]The University of Tokyo, Kashiwa 277-8561, Japan

E-mail: bando.takahiro@qst.go.jp

**Abstract**

In JT-60U, simultaneous excitation of $n = 1$ helical cores (HCs) and $m/n = 2/1$ Tearing Modes (TMs) was observed [T. Bando et al., *Plasma Phys. Control. Fusion* **61** 115014 (2019)]. In this paper, we have investigated the excitation mechanism of $n = 1$ HCs with $m/n = 2/1$ TMs based on the experimental observations and a simple quasi-linear MHD model. In the previous study, it was reported that a "coupling" on the phase of the MHD mode is observed between $n = 1$ HCs and $m/n = 2/1$ TMs: (1) The $m/n = 2/1$ TM and the $n = 1$ HC rotate keeping $f_{m/n=1/1(HC)} = 2f_{m/n=2/1(TM)}$, resulting in the phase locking. Here, $f_{m/n=1/1(HC)}$ and $f_{m/n=2/1(TM)}$ are the rotation frequency of the $m/n = 1/1$ HC and the $m/n = 2/1$ TM in the poloidal direction, respectively. (2) The core is shifted to the high-field side when the O-points of the $m/n = 2/1$ magnetic island line up in the midplane. In this study, it is found that the coupling is observed with the mode frequency from several Hz to 6 kHz. This indicates that the resistive wall and the plasma control system do not induce the coupling because the both time scales are different from the mode frequency. In addition, $n = 1$ HCs appear to be the non-resonant mode from the two observations: $n = 1$ HCs do not rotate with the plasma around the $q = 1$ surface in the core and the coupling is also observed even when $q_{min} > 1$. It is also observed that the electron fluctuation due to an $n = 1$ HC in the core region disappears with the stabilization of an $m/n = 2/1$ neoclassical tearing mode by electron cyclotron current drive, implying that $n = 1$ HCs are driven by $m/n = 2/1$ TMs. This perspective, $n = 1$ HCs are driven by $m/n = 2/1$ TMs, is supported by the observation that the saturated amplitude of the $m/n = 1/1$ component of the radial displacement in the core is smaller than that of the $m/n = 2/1$ component. Finally, we revisit a quasi-linear MHD model where the $m/n = 1/1$ HC is induced directly by the sideband of the current for the $m/n = 2/1$ TM, which allows to excite the non-resonant $m/n = 1/1$ mode. The model also describes the characteristic of the coupling, $f_{m/n=1/1(HC)} = 2f_{m/n=2/1(TM)}$.

## 1. Introduction

Helical cores (HCs) in tokamaks have been an important research topic for the fusion reactor because HCs can affect transports of impurities [1-4], energetic particles [5], or toroidal momentum [5]. In addition to the effect on transport by helical structures, recent studies suggested that magnetohydrodynamics (MHD) dynamo accompanied with HCs redistributes the current profile in the core and realizes sawtooth-free plasmas of hybrid scenarios as observed in DIII-D [6,7], in which the minimum value of the safety factor profile, $q_{min.}$, is kept slightly above unity by "flux pumping".

Several theoretical models to explain the excitation mechanism of HCs have been proposed. The generation of the magnetic island by the excitation of resonant tearing modes (TMs) due to the decrease of the electron temperature around the $q = 1$ surface was proposed [8], which is induced by hydrogen pellet injection. This model intended to explain "snake" structure in soft X-ray (SX) emission, which was observed in JET [9]. HC equilibria can be obtained with $q_{min.} \sim 1$ utilizing a three-dimensional MHD equilibrium solver [10,11], which is consistent with experimentally observed HCs in plasmas having $q_{min.} \sim 1$ [4,12,13,14]. The robustness of HC equilibria is confirmed by the linear and nonlinear MHD stability calculations [15]. If the beta value is sufficiently high, the saturated $m/n = 1/1$ quasi-interchange instability is excited and induce the dynamo loop voltage using 3D nonlinear simulations [16]. On the other hand, in recent experimental studies, HCs are induced by $n = 1$ magnetic perturbation with the external coils and characteristics of HCs are investigated with taking account of reconstructed three-dimensional MHD equilibria in DIII-D [7,17,18]. The EAST team recently showed that the two-dimensional mode structure of a long-lived helical mode by combined use of tomography and a singular value decomposition, showing an $m/n = 1/1$ mode structure and a substantial $m/n = 2/2$ component where $q_{min}$ is kept slightly above unity [19].

Though many experimental and theoretical studies have been reported as mentioned above, the study on the excitation mechanism of HCs with MHD modes except for $m/n = 1/1$ modes has not been reported as far as we know. Recent our study [20] showed that a "coupling" on the phase of the MHD mode is observed between $n = 1$ HCs and $m/n = 2/1$ TMs: (1) The $m/n = 2/1$ TM and the $m/n = 1/1$ HC rotate keeping $f_{m/n=1/1(HC)} = 2f_{m/n=2/1(TM)}$, resulting in the phase locking. Here, $f_{m/n=1/1(HC)}$ and $f_{m/n=2/1(TM)}$ are the rotation frequency of the $m/n = 1/1$ HC and the $m/n = 2/1$ TM in the poloidal direction, respectively. (2) The core is shifted to the high-field side when the O-points of the $m/n = 2/1$ magnetic island line up in the midplane as shown in figure 1. In this study, the "coupling" means coupling satisfying the two characteristics (1) ~ (2) mentioned in this paragraph. In our previous study [20], though possible excitation mechanisms of HCs were proposed such as the effect of the resistive wall where the resistive wall time $\tau_w$ ~ 10 ms [21] in JT-60U or helical equilibria with $q_{min}$ ~ 1, these candidates were not investigated. This is mainly because the coupling was only studied with TMs having low mode frequencies (<= 20 Hz), mentioned as Low Frequency Mode (LFM) in the previous study [20]. In addition, it is also required to investigate whether the magnetic fluctuation induced by the plasma control system, where the time interval of the control of the plasma position is 250 µs (4kHz), can induce HCs or not. This is because HCs induced by the external $n = 1$ component of the magnetic fluctuation are observed in DIII-D and ASDEX [4]. Though the dominant component of the magnetic fluctuation induced by the plasma control system is the $n = 0$ component, the $n = 1$ component can be induced due to the misalignment of the coils in JT-60U.

In this paper, we report investigations of the excitation mechanism of $n = 1$ HCs in JT-60U. It is found that the mode frequency with the coupling can be 6 kHz, which indicates that the resistive wall and the plasma control system do not induce the coupling. In addition, $n = 1$ HCs appear to be the non-resonant mode from two observations. First, it is found that $m/n = 1/1$ HCs do not rotate with the plasma around the $q = 1$ surface in the core region. Second, the coupling is observed even when $q_{min} > 1$. It is also observed that the temperature fluctuation due to an $n = 1$ HC in the core region disappears with the stabilization of an $m/n = 2/1$ neoclassical tearing mode (NTM), implying that $n = 1$ HCs are driven by $m/n = 2/1$ TMs. This perspective, $n = 1$ HCs are driven by $m/n = 2/1$ TMs, is supported by the observation that the saturated amplitude of the $m/n = 1/1$ component of the radial displacement is smaller than that of the $m/n = 2/1$ component. We revisit a quasi-linear MHD model where the $m/n = 1/1$ HC is induced directly by the sideband of the current for the $m/n = 2/1$ TM, which allows to excite the non-resonant $m/n = 1/1$ mode. The model consistently describes the characteristic of the coupling, $f_{m/n=1/1(HC)} = 2f_{m/n=2/1(TM)}$.

This article is organized as follows: In section 2, the heating system and diagnostic of JT-60U are introduced. In section 3, the investigations of $m/n = 1/1$ HCs with $m/n = 2/1$ NTMs with the mode frequencies of several kHz are introduced in subsection 3.1 first, then, the observation of an $m/n = 1/1$ HC, where $q_{min}$ is greater than unity, is introduced in subsection 3.2. In these subsections, the observed mode structures are compared with the characteristics of the coupling mentioned in this section. The causality between the $m/n = 1/1$ HC and the $m/n = 2/1$ TM is investigated in section 4. The discussion focusing on the excitation mechanism is given in section 5. We revisit the quasi-linear MHD model where the $m/n = 1/1$ mode is induced by the sideband. Then the study is summarized in section 6.

## 2. Heating system and diagnostic of JT-60U

JT-60U is a tokamak device where typical major radius is 3.4 m and typical minor radius is 0.9 m [22]. The toroidal direction of the toroidal magnetic field and the plasma current is the clockwise direction in the analyzed discharges. In JT-60U, plasmas are heated by eleven neutral beam injectors (NBIs) with positive-ion source (P-NBI), whose acceleration energies are about 85 keV. Two (#9 and #10) of the tangential P-NBIs are co-current direction (co-direction) and other two (#7 and #8) tangential P-NBIs are counter-current direction (ctr-direction). Here, the co-direction and the ctr-direction are defined as the same direction and the opposite direction to the plasma current, respectively. The other seven NBs inject neutral beams to the plasma perpendicularly (PERP P-NBI). The electron cyclotron heating (ECH) with gyrotron is also available. The ECH is used to stabilize NTMs by electron cyclotron current drive (ECCD) [23] as well as to heat plasmas.

Figure 2 shows the sight lines of SX measurement (Upper SX) and the measured plane by electron cyclotron emission (ECE) measurement by heterodyne radiometers [24] and the polychromator [25], which are used to study the radial mode structure. The MHD equilibrium in figure 2 is obtained at 9.5 s of 49713 (figures 3, 4, and 5). The difference in the toroidal angle of the locations of SX measurement and ECE measurement are about 160 degrees. The ion temperature and the toroidal velocity are measured by charge exchange recombination spectroscopy (CXRS) [26], which is the same as CXRS1 in [20]. The toroidal and poloidal mode numbers are estimated with magnetic probes arranged toroidally and poloidally. The profile of the safety factor ($q$) is calculated with 16 channels of the motional Stark effect (MSE) measurement [27,28] on the P17 port of JT-60U, which measures the polarization angles from the core region to the edge region of plasmas. The $q$-profile is reconstructed with the MEUDAS code [29] to reconstruct MHD equilibria assuming the toroidally axisymmetric geometry by solving the Grad-Shafranov equation. The MHD equilibrium is reconstructed so that the $q$-profile from the MHD equilibrium is consistent with the $q$-profile from the polarization angle of MSE measurement in the P17 port. The reconstructed MHD equilibrium can be obtained every 0.01 s, which is determined by the time resolution of the polarization angle from MSE measurement.

## 3. Experimental condition of observation of m/n = 1 HC with m/n = 2/1 TM

### 3.1 m/n = 1 HC with m/n = 2/1 NTM whose frequency is several kHz

$n = 1$ HCs with $m/n = 2/1$ NTMs, whose observed frequencies are several kHz, are investigated in the stabilization experiments of $m/n = 2/1$ NTMs by ECCD [23]. The plasma parameters in the analyzed plasmas are shown in table 1. In these discharges, NTMs are excited by the high injection power (~ 20 MW) of NBI and then the mode frequency of NTMs is increased with the increased toroidal rotation velocity by the tangential NBIs. As results, the mode frequencies of NTMs become several kHz. Due to the large net toroidal torque to the ctr-direction, the rotation directions of NTMs are the ctr-direction toroidally and the electron diamagnetic direction poloidally in discharges of table 1 when the mode frequencies of $m/n = 2/1$ NTMs are several kHz.

In the analyzed discharges, the coherent fluctuations are observed in SX signals and the electron temperature in ECE measurement with the mode frequency of the $m/n = 2/1$ NTM. Because the observed mode frequencies are the same, the phase profile can be evaluated in both measurements. The phase profiles are calculated at three times in the discharge of figure 3. In the time window of figure 3, $\beta_p$ (an indicator of the saturated size of the magnetic island of NTMs [32]) and the total injection power of NBIs are nearly constant. As shown in figure 4 (a) and (b), the unchanged phase profiles of $m = 1$ structure in SX signals are observed, indicating the rotating $m = 1$ structure. Note that, in this study, the $m = 1$ structure is not measured with magnetic probes arranged poloidally. This might be due to the small amplitude of the $m = 1$ component compared with the $m = 2$ component. In the case of ECE measurement, the change of the phase is just $0.14\pi$ rad. across $\rho = 0$ and is significant in the high field side where the sign of $\rho$ is minus. Here, $\rho$ is the volume-averaged minor radius. This is because the difference of the poloidal angles of the measured points across $\rho = 0$ is not $\pi$ rad. on the horizontal plane for ECE measurement as seen in figure 2. In figure 4 (c), the estimated phases projected on the midplane are shown assuming the measured points of ECE measurement are moved to the midplane keeping the measured minor radius. As seen in the estimated phase profile of figure 4 (c), the difference of the phase across $\rho = 0$ is about $\pi$ rad., indicating $m = 1$ structure. The detailed description of the modification is found in Appendix A. The $m = 1$ structure in SX measurement and ECE measurement seems to have $n = 1$ structure because the relationship between the phase profiles around $\rho = 0$ in SX measurement and ECE measurement (figure 4 (a) and (b)) can be reproduced in synthetic images assuming the existence of the $m/n = 1/1$ mode structure in the core, which is the same method as a method in the previous study [20]. The phase difference $\pi$ around the $q = 2$ surface of figure 4 (d) in ECE signals is due to the $m/n = 2/1$ NTM. Therefore, the unchanged phase profiles of $m/n = 1/1$ component and $m/n = 2/1$ component are obtained, indicating phase locking. The observed phase locking is corresponding to the characteristics (1) and (2) of the coupling because the observed phase profiles can be reproduced in synthetic images assuming the coupling. Note that the same phase from $\rho = 0.1$ to $\rho = 0.5$ in ECE measurement is considered due to the $m = 1$ "helical deformation" mentioned in the previous study [20].

Because the observed mode frequency with the coupling can be several kHz in this study as well as ~ 20 Hz in the previous study [20], the resistive wall does not induce the coupling. In addition, the coupling is observed when the mode frequency of an $m/n = 2/1$ NTM is 6 kHz. This indicates that the observed HCs are not induced by the plasma control system of JT-60U. Because the plasma control system controls the position of the plasma with 4 kHz by changing the coil current, the fluctuation having the frequency with 6 kHz cannot be induced.

As indicated in figure 4, the $n = 1$ HC and the $m/n = 2/1$ NTM have the same mode frequency. If the observed $n = 1$ HC rotates with the bulk plasma like NTMs, the relationship, $f_{mode} \sim n \times f_{tor} - m \times f_{pol}$, should be satisfied [33]. Here, $f_{mode}$ is the "observed" mode frequency, $f_{tor}$ is the toroidal rotation frequency defined by $V_{tor}/(2\pi R_m)$, and $f_{pol}$ is the poloidal rotation frequency defined by $V_{pol}/(2\pi \rho_m)$. Here, $V_{tor}$ ($V_{pol}$), $R_m$, and $\rho_m$ are the toroidal (poloidal) rotation velocity, the major radius where $V_{tor}$ is measured, and the minor radius where $V_{pol}$ is measured, respectively. In this study, $f_{pol}$ is ignored because the net toroidal torque induced by the tangential NBIs are applied and $V_{tor}$ is sufficiently larger than $V_{pol}$ estimated with the CHARROT code [34,35] assuming the effective ion charge $Z_{eff}$ is 2. The absolute values of the poloidal flow of the bulk deuterium plasma are smaller than 0.6 km/s from core region to the edge region. Figure 5 shows the radial profile of (a) $V_{tor}$ measured by CXRS measurement and (b) $f_{tor}$ at 9.5 s of 49713 in figure 3, assuming that the rotation velocity of the bulk deuterium plasma and the carbon plasma as impurity are the same. The purple dashed line in figure 5 (a) is the calculated toroidal rotation velocity profile of the deuterium by the CHARROT code. The profile of the toroidal rotation velocity by CXRS measurement and the profile by the CHARROT code are consistent, indicating that it is valid to use $V_{tor}$ of carbon plasma as the rotation velocity of the bulk deuterium plasma. The mode frequency of the $m/n = 2/1$ NTM in figure 3 (b) is about 2.3 kHz and is consistent with $f_{tor}$ around the $q = 2$ surface as seen in figure 5 (b). On the other hand, if the $m/n = 1/1$ HC rotates with the plasma around the $q = 1$ surface, the observed frequency should be ~ 5 kHz from $f_{tor}$ around $q = 1$ surface ($\rho \sim 0.2$) in figure 5(b) (see also $q$-profile in figure 4 (d)). However, the experimentally observed frequency of the $n = 1$ HC is the same as that of the $m/n = 2/1$ NTM. This indicates that the observed $m/n = 1/1$ HC cannot be explained by the rotation at the $q = 1$ surface located at the core region in the plasma. In other words, there is a possibility that the observed $n = 1$ HC is a non-resonant mode.

## 3.2. m/n = 1 HC with m/n = 2/1 TM with $q_{min}$ > 1

As mentioned in subsection 3.1, the observed $n = 1$ HC is possibly a non-resonant mode. However, in figure 4 (d), the $q$-profile is the normal shear where $q_{min}$ is nearly unity. Therefore, it is better to examine whether the $m/n = 1/1$ structure is observed with $q_{min} \gg 1$ or not. This is examined at 7.55 s in 48885 where an $m/n = 2/1$ TM is observed. The plasma discharge 48885 is for a high β plasma experiment [36] and it is tried to make weak shear profile for forming the internal transport barrier (ITB). With the weak shear profile, $q_{min}$ tends to greater than unity.

The extended view of 48885 is shown in figure 6 (a) ~ (d). In figure 6 (a), the $n = 1$ mode, whose mode frequency is about 200 Hz, is observed in the magnetic fluctuation before 7.55 s. The mode frequency of the $m/n = 2/1$ TM is relatively slow (~ Hz) after the decrease of the mode frequency at 7.55 s, which is the same object as the LFM described in the previous study [20]. And in figure 6 (b) ~ (d), the coherent fluctuations are observed in SX signals and the electron temperature with the mode frequency of the $n = 1$ mode. In figure 6 (b), because the $q = 2$ surface locates around ρ = 0.6 and the phases of the fluctuations are inverted between ρ ~ 0.50 and ρ ~ 0.57, it is confirmed that the observed $n = 1$ mode is an $m/n = 2/1$ TM. As well as the phases around $q = 2$ surface, the phases are inverted between signals across ρ = 0 in figure 6 (c) and (d), indicating the $m = 1$ structure. The $m = 1$ structure in SX measurement and ECE measurement seems to have $n = 1$ structure because the relationship between the phase profiles across ρ = 0 in SX measurement and ECE measurement (figure 6 (c) and (d)) can be reproduced in synthetic images assuming the existence of the $m/n = 1/1$ mode structure in the core which is the same as the result in the previous subsection. As seen in figure 6 (a) ~ (d), the phases are locked (phase locking). Because the observed phases can be reproduced in synthetic images assuming the coupling, the characteristics (1) and (2) of the coupling are confirmed in this discharge as well.

The $q$-profile estimated with the MEUDAS code is shown in figure 6 (e). In this estimation, the polarization angle is averaged from 7.54 s to 7.56 s for the reconstruction of the MHD equilibria to average the non-axisymmetric effect of the mode structure. As shown in figure 6 (e), $q_{min}$ at 7.55 s is about 1.4 > 1, namely $q_{min}$ > 1, indicating that the observed $m/n = 1/1$ mode is a non-resonant mode.

## 4. Causality between m/n = 1/1 HC and m/n = 2/1 TM

In the previous section, it is found that $n = 1$ non-resonant HCs and $m/n = 2/1$ TMs are coupled. In this section, the causality between $m/n = 1/1$ HCs and $m/n = 2/1$ TMs is investigated to consider the excitation mechanism of the $n = 1$ HC and the coupling.

As shown in figure 6 (e), the $n = 1$ HC is observed with the low magnetic shear. With a low magnetic shear, it was reported that the $m/n = 1/1$ infernal mode can induce the $m/n = 2/1$ NTM [37,38]. In this study, however, it is not observed that an $n = 1$ HC firstly appears and an $m/n = 2/1$ NTM subsequently appears excluding the case with the sawtooth collapse, which occasionaly induces NTMs [39]. Therefore, the possibility that the $m/n = 2/1$ NTM induces the $n = 1$ HC is investigated in this study.

Here, the fluctuation in the core and the fluctuation around the $q = 2$ surface are investigated when an $m/n = 2/1$ NTM is stabilized by ECCD around the $q = 2$ surface in 49573 as shown in figure 7, which is one of the series of discharges in table 1. As seen in figure 7, the magnetic fluctuation and the temperature fluctuation by the $m/n = 2/1$ NTM on the $q = 2$ surface (blue) is stabilized by ECCD around the $q = 2$ surface until 11 s as indicated by the vertical dashed red line. As seen in figure 7 (b), the temperature fluctuation by the $n = 1$ HC from core (red) also disappears until the vertical red line. Because the temperature fluctuation in the core region due to the $n = 1$ HC disappears with the stabilization of the $m/n = 2/1$ NTM at the $q = 2$ surface, it is natural to consider that the $m/n = 1/1$ HC is induced by the $m/n = 2/1$ NTM.

This perspective, $n = 1$ HCs are driven by $m/n = 2/1$ TMs, is supported by observation of the radial displacement ξr with the coupling of an $m/n = 1/1$ HC and an $m/n = 2/1$ TM in figure 8. As shown in figure 8, ξr is about 1 ~ 1.5 cm in the core region, which seems to be the $m/n = 1/1$ component of ξr, (ξr)$_{1,1}$. On the other hand, the $m/n = 2/1$ component of ξr around the $q = 2$ surface, (ξr)$_{2,1}$, is much larger than (ξr)$_{1,1}$ in the core region. The larger (ξr)$_{2,1}$ indicates that the driving source of the $m/n = 1/1$ HC is the $m/n = 2/1$ TM.

## 5. Discussion

As discussed in subsection 3.1, the same observed frequencies of $m/n = 2/1$ NTMs and $n = 1$ HCs cannot be explained by rigid plasma rotation. Here, if the $n = 1$ HC is the sideband of the $m/n = 2/1$ TM, the same observed frequency and the characteristic (1) of the coupling can be explained. However, the way that the $m/n=2/1$ TM drives the spatially separated $n = 1$ HC in the core region is non-trivial. Here we propose a classical forced magnetic reconnection (FMR) model [40,41] based on the quasi-linear MHD model to explain the non-resonant mode and the characteristic of the coupling (1). As discussed in section 4 on the causality, the $m/n = 2/1$ TM is induced firstly and the $n = 1$ HC is induced subsequently. If the $m/n = 2/1$ TM exists, the $m/n = 2/1$ current filament can induce the $m/n = 1/1$ mode as the sideband at "$q = 2$" rational surface, not at the "core". Once the $m/n = 1/1$ mode is excited at the "$q = 2$" surface, the $m/n = 1/1$ mode propagates into the core with Alfvenic speed [40] until causing the discontinuity at the rational surface, or surviving as an Alfven eigen modes (non-resonant mode). The former discontinuity causes the FMR and also forms the $m/n = 1$ mode at the core. Thus, from the FMR point of view, the $m/n = 2/1$ TM can drive the spatially separated $n = 1$ HC regardless of the resonant or the non-resonant mode. The schematic view of the excitation mechanism is shown in figure 9. Regarding the phase relation, metrics are all even under 1st order large aspect ratio approximation. Therefore, the phase of the $n = 1$ mode at the $q = 2$ surface is the same with that of the $m/n = 2/1$ TM. The $m/n = 1/1$ mode at the $q = 2$ surface is the boundary condition for the $m/n = 1/1$ mode at the core. Thus, the $m/n = 1$ mode is in-phase with the $m/n = 2/1$ TM, which is not consistent with the observation, while it successfully explains the two modes have the same observed frequency and the characteristic (1) of the coupling, $f_{m/n=1/1(HC)} = 2f_{m/n=2/1(TM)}$. The discrepancy may be explained by the theoretical explanation of the phase locking with the dynamical stability in [42] and its experimental validation in [43]. The further study is required on this issue.

Though the revisited quasi-linear MHD model can explain the non-resonant characteristic and the characteristic (1) of the coupling, there are some issues to deal with in future works. For example, the revisited model is just for the linear phase and the observation of the nonlinear phase, such as the radial displacement in figure 8, cannot be treated. In addition, the characteristic (2) of the coupling and "helical deformation" mentioned in [20] should be also explained. The nonlinear simulations are required to investigate these issues.

In our previous study [20], the similarity between quasi-stationary modes (QSMs) in the JET tokamak [44,45] and our observations is discussed. QSMs are the low frequency modes (~ 10 Hz) and showing the $m = 1$ structure with $m/n = 2/1$ modes, which is the same as $n = 1$ HCs with $m/n = 2/1$ TMs in our study. Though QSMs were only observed with ~ 10 Hz, we show the coupling with the higher observed frequency with several kHz and introduce the possibility of the non-resonant $m = 1$ mode in the core. In consideration of the similarity between QSMs and $n = 1$ HCs in our study, there is possibility that QSMs are accompanied by the non-resonant mode and this perspective should be examined.

## 6. Summary and conclusion remarks

In summary, in this paper, we investigate the excitation mechanism of $n = 1$ HCs coupled with $m/n = 2/1$ TMs described in the previous study [20]. It is found that the coupling is observed with the mode frequency from several Hz to 6 kHz. This indicates that the resistive wall and the plasma control system do not induce the coupling. In addition, $n = 1$ HCs are found to be the non-resonant mode from two observations. First, it is observed that $m/n = 1/1$ HCs do not rotate with the plasma around the $q = 1$ surface in the core region by means of the comparison of the plasma rotation frequency $f_{tor}$ and the observed mode frequency $f_{mode}$. Second, the coupling is observed even when $q_{min} > 1$. Then the causality between $m/n = 1/1$ HCs and $m/n = 2/1$ TMs is investigated with the observation of the disappearance of the electron fluctuation in the core with the stabilization of an $m/n = 2/1$ NTM, indicating that $n = 1$ HCs are induced by $m/n = 2/1$ TMs. We revisit a traditional FMR based on the quasi-linear MHD model where the $m/n = 1/1$ HC is induced directly by the sideband of the current for the $m/n = 2/1$ TM. The model can explain the non-resonant characteristic and the characteristic of the coupling, $f_{m/n=1/1(HC)} = 2f_{m/n=2/1(TM)}$. Nonlinear MHD simulations are still required to explain the characteristics in the nonlinear phase observed at experiments in future.

In this study, we report observations of non-resonant $m/n = 1/1$ HCs coupled with $m/n = 2/1$ TMs for the first time and revisit the traditional quasi-linear MHD model where the $m/n = 1/1$ mode is induced as the sideband of $m/n = 2/1$ mode to explain the observed characteristics. In other tokamaks [4] or an RFP [46], it is observed that the external magnetic perturbations induced the helical structure in the plasma. If the $n = 1$ HC is surely induced by the $m/n = 2/1$ NTM as proposed in this study, it can be considered that the $m/n = 2/1$ NTM plays the role of the external magnetic perturbations to indue the helical structure with the same mechanism in other toroidal devises. This also should be examined theoretically and experimentally.

**Acknowledgments**

Helpful comments for analysis by Dr T. Suzuki, Dr. N. Aiba and Dr. S. Sumida are greatly appreciated.

## Appendix A: Method to modify the phase in subsection 3.1

If we distinguish the poloidal number in the core, the change of the poloidal angle of the measured points should be around π rad. across ρ = 0. However, as shown in figure 10 (a), the poloidal angles of ECE measurement in the core region changes 0.7 π rad. Therefore, it is required to modify the observed phase considering these poloidal angles to investigate the mode structure. Here, as shown in figure 10 (b), if the measured point before the movement is located as the lower filed side from the magnetic axis, the measured point is moved to the point on the midplane in the lower field side keeping the minor radius. In the opposite case, the measured position is moved to the higher field side. This method to modify the phase is valid when the single mode having a certain poloidal number covers the region for the estimation. This method is also affected by the accuracy of the estimation of the MHD equilibrium. The error due to the estimation of the MHD equilibrium is more sensitive near the estimated axis.

| | |
|---|---|
| $\beta_N$ | 0.6 ~ 0.8 % m T MA$^{-1}$ |
| $\beta_p$ | 0.46 ~ 0.68 |
| Total neutron emission rate | 1.1x10$^{15}$ ~ 2.5x10$^{15}$ n/s |
| Plasma current | 1.5 MA |
| Magnitude of magnetic field at magnetic axis | 3.7 T |
| Major radius | 3.18 m |
| Minor radius | 0.8 m |
| Plasma volume | 55 ~ 56 m$^3$ |
| Radial position of axis | 3.23 ~ 3.25 m |
| Vertical position of axis | 0.10 ~ 0.11 m |
| Plasma triangularity | 0.18 ~ 0.20 |
| Plasma elongation | 1.4 ~ 1.5 |
| q$_{95}$ (the safety factor on 95% of flux surface) | 4.05 ~ 4.09 |
| Plasma-wall separation in low field side | 0.58 ~ 0.60 m |

Table 1: The plasma parameters of 39 discharges where $m/n = 2/1$ NTMs with $n = 1$ HCs, whose mode frequencies are several kHz, are observed. The analyzed discharges are made after the installation of ferrite steel tiles on the vacuum vessel to reduce the toroidal magnetic field ripple [30].

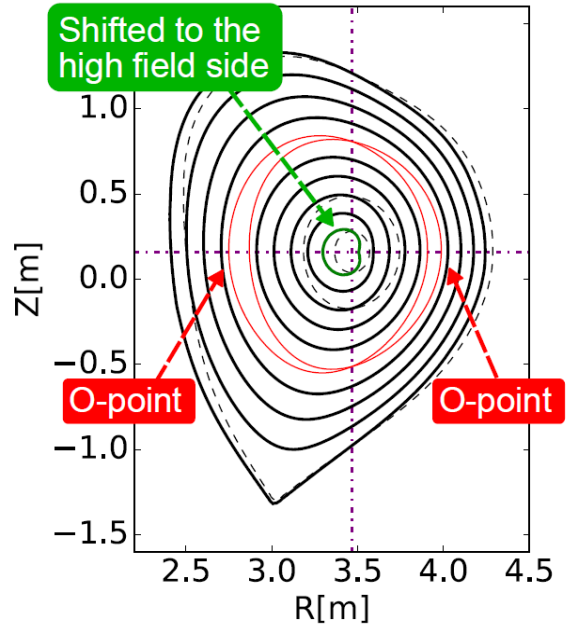

Figure. 1. Schematic view of the poloidal cross sections showing the phases of the $m/n = 2/1$ TM and the $m = 1$ HC based on the view of the coupling described in the previous study [20]. The core is shifted to the high-field side when the O-points of the $m/n = 2/1$ magnetic island line up in the midplane. To keep this configuration, the $m/n = 2/1$ TM and the $n = 1$ HC have to rotate satisfying $f_{m/n=1/1(HC)} = 2f_{m/n=2/1(TM)}$. The red curves represent the $m = 2$ magnetic island. The black and green curves show the flux surfaces with $m = 1$ deformation. The black dashed lines show the flux surfaces at $\rho \sim 0.1, 0.3,$ and 1.0 without $m = 1$ deformation. The cross points of the purple dashed lines show the magnetic axis without $m = 1$ deformation. The deformed flux surfaces are obtained by adding artificial $m/n = 1/1$ deformation and artificial $m/n = 2/1$ magnetic islands to the axisymmetric equilibrium.

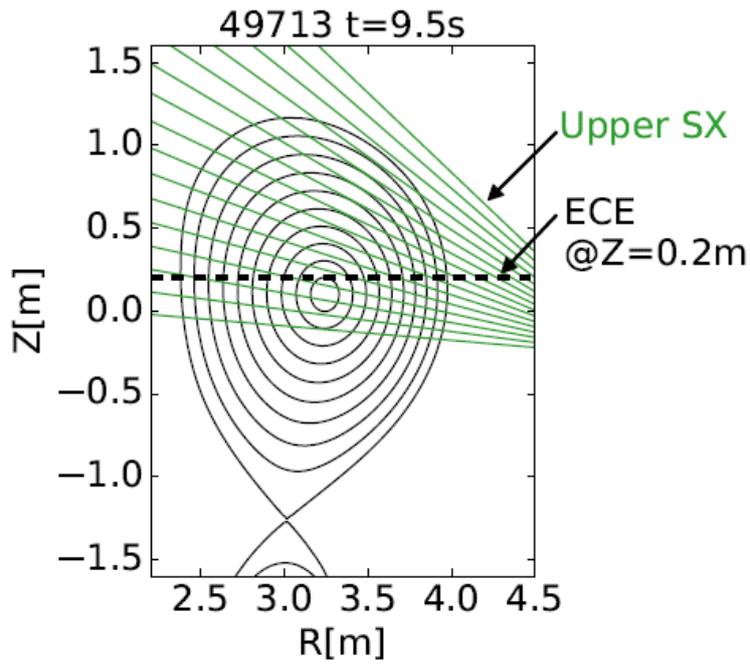

Figure. 2. The sight lines of SX measurement and the measured plane by ECE measurement. The difference in the toroidal angles between the locations of SX measurement and ECE measurement are about 160 degrees. The MHD equilibrium is obtained at 9.5s of 49713.

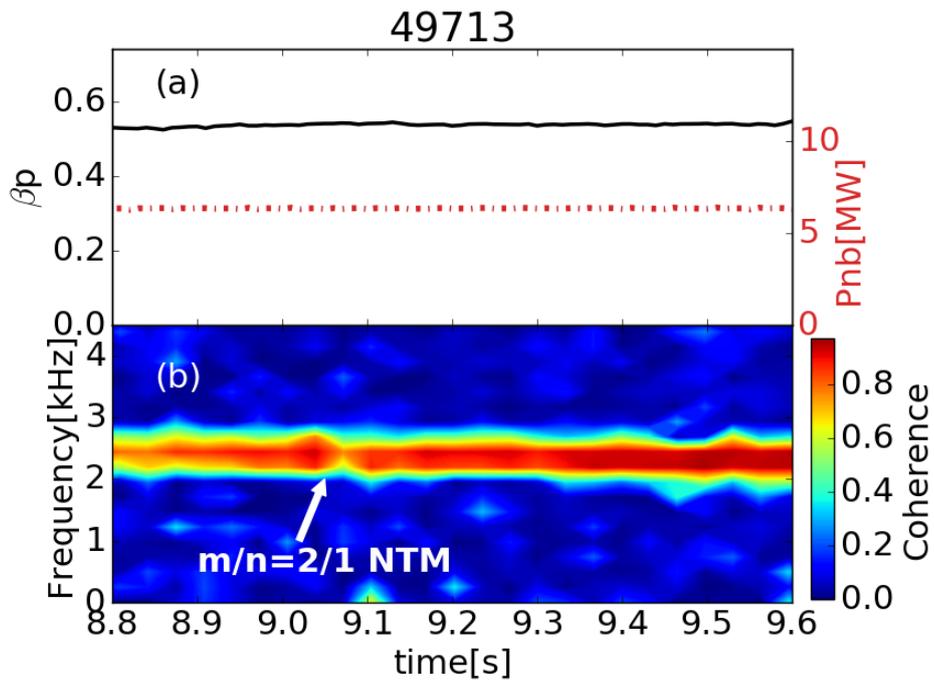

Figure. 3. Time evolution of the signals from an experiment of the stabilization of *m/n* = 2/1 NTMs by ECCD in JT-60U: (a) the poloidal beta and the total injection power of NBIs (dotted-dashed red line) and (b) the spectrum of coherence of SX signals from a slight line viewing the core and the magnetic fluctuation.

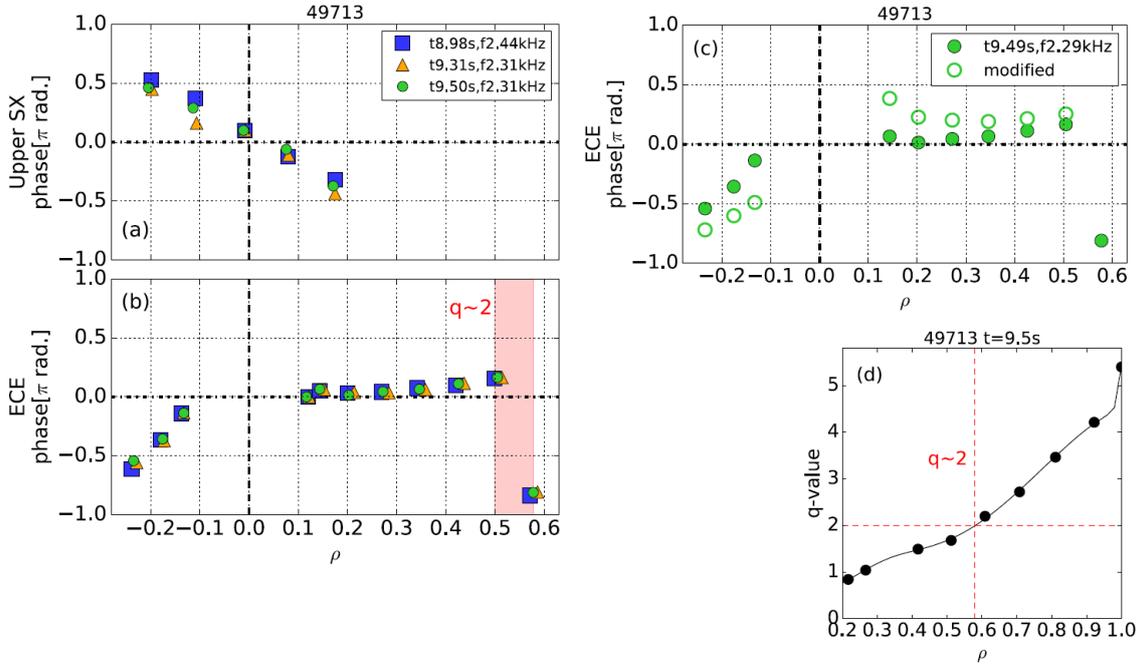

Figure. 4. The radial profile of the phase obtained with cross spectrum of Fourier transform at 9 s, 9.3 s, and 9.5 s from (a) SX measurement and (b) ECE measurement. (c) The radial profile of the phase and the modified phase at 9.5 s. (d) The $q$-profile at 9.5 s obtained with MSE measurement and the MEUDAS code. In (a), (b), and (c), the coherence is calculated between the channel viewing near the axis and other channel. The phases where the coherence is high are shown. In (c), the data with the poloidal angle ~ $0.5 \pi$ rad. is not shown because the error of the modification near the axis can be large as explained in Appendix A. In (d), the black circles are the estimated safety factor of MSE channels. Because the measure points do not exist in the core region $\rho < 0.215$, the $q$-profile is not shown in the core region.

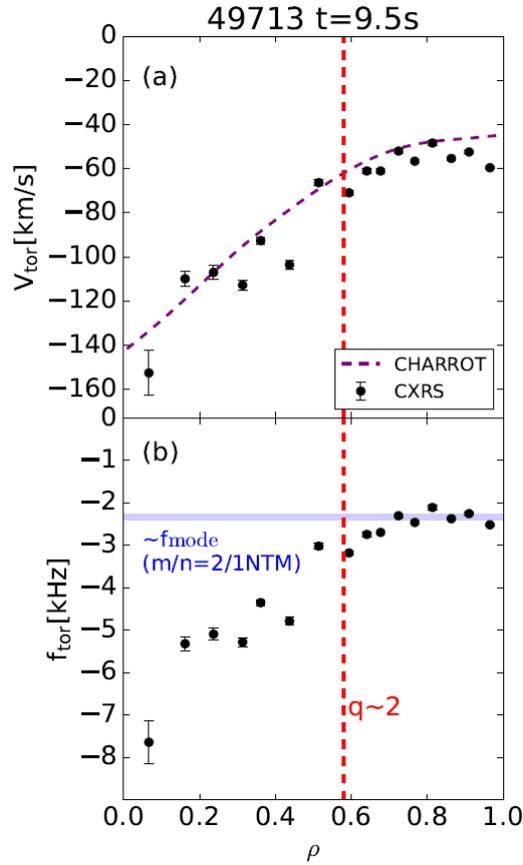

Figure. 5. The radial profile of the toroidal velocity ($V_{tor}$) and the toroidal rotation frequency ($f_{tor}=V_{tor}/2\pi R_m$) measured by CXRS measurement at 9.5 s. Here, $R_m$ is the major radius where $V_{tor}$ is measured. In (a), the purple dashed line is the toroidal velocity of bulk deuterium plasma estimated by the CHARROT code [34,35]. In (b), $f_{mode}$ is the observed mode frequency of the $m/n = 2/1$ NTM. Here, the monotonic increase in the toroidal velocity profile is due to the net toroidal torque to the ctr-direction from the tangential NBI.

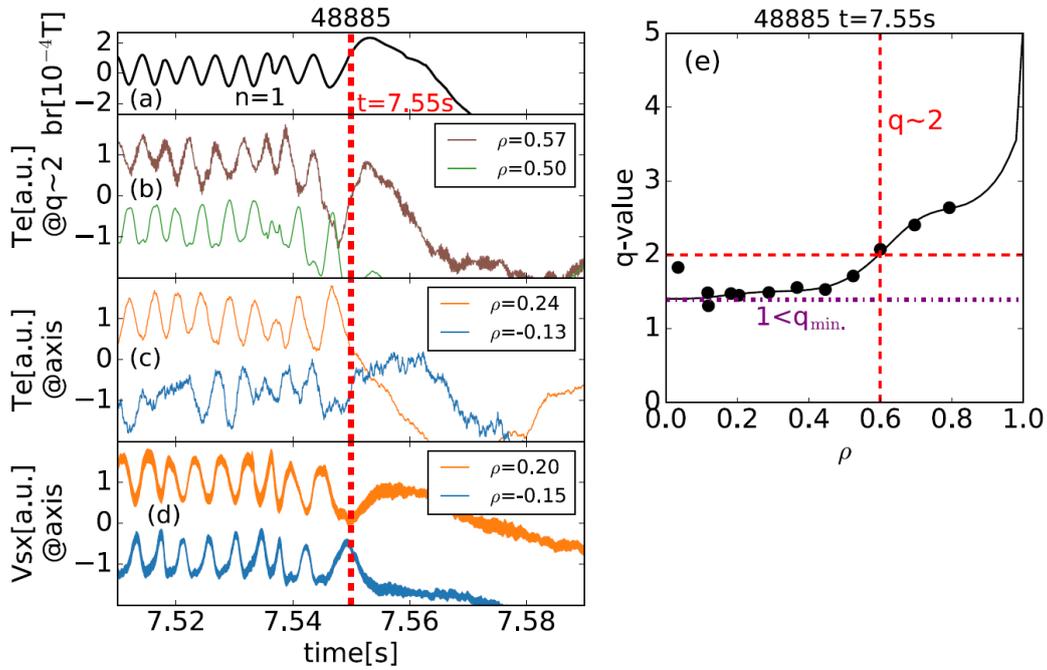

Figure. 6. The time evolution of (a) the magnetic fluctuation, (b) the electron temperature at $\rho = 0.50$ and $0.57$ around the $q = 2$ surface, (c) the electron temperature at $\rho = -0.13$ and $0.24$, and (d) SX signals from $\rho = -0.15$ and $0.2$. (e) The $q$-profile at 7.55 s estimated with MSE measurement and the MEUDAS code. In (e), the black circles are the estimated safety factor of MSE channels. At 7.55 s, the plasma parameters are the plasma current ~ 0.9 MA, the magnitude of magnetic field at magnetic axis ~ 1.52 T, the major radius ~ 3.43 m, the minor radius ~ 0.92 m, the radial position of axis ~ 3.53 m, the vertical position of axis ~ 0.2 m, the plasma elongation ~ 1.39, the plasma triangularity ~ 0.38, the plasma volume ~ 72 m$^3$, and $q_{95}$ ~ 3.5. This discharge is made after the installation of ferrite steel tiles on the vacuum vessel to reduce the toroidal magnetic field ripple [29]. Here, the difference in the poloidal angle between the measured points across $\rho \sim 0$ in ECE measurement are about $\pi$ rad. unlike the case in figure 4 (b) because the vertical position of axis and the measured plane by ECE measurement are on the same horizontal plane.

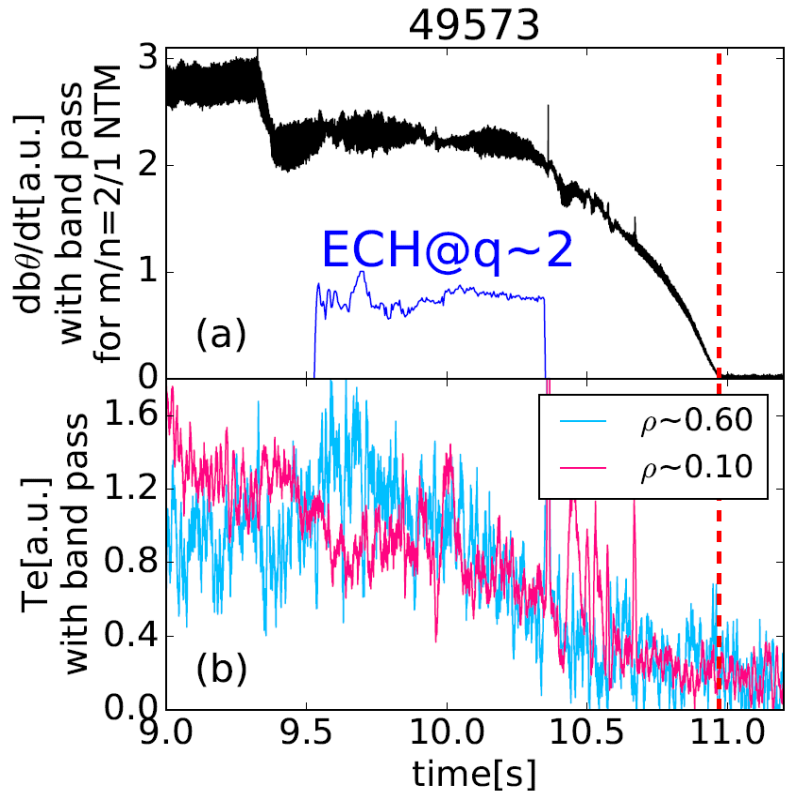

Figure.7. The time evolution of (a) the magnetic fluctuation and (b) amplitudes of signals from ECE measurement with the band pass filter to obtain the component due to an $m/n = 2/1$ NTM. In (a), the blue line indicates the pulse pattern of ECH (ECCD) around $q = 2$ surface to stabilize the $m/n = 2/1$ NTM. The $m/n = 2/1$ NTM is stabilized around 11 s according to the magnetic fluctuation. In (b), the signals measured at $\rho = 0.108$ and $0.602$ are shown.

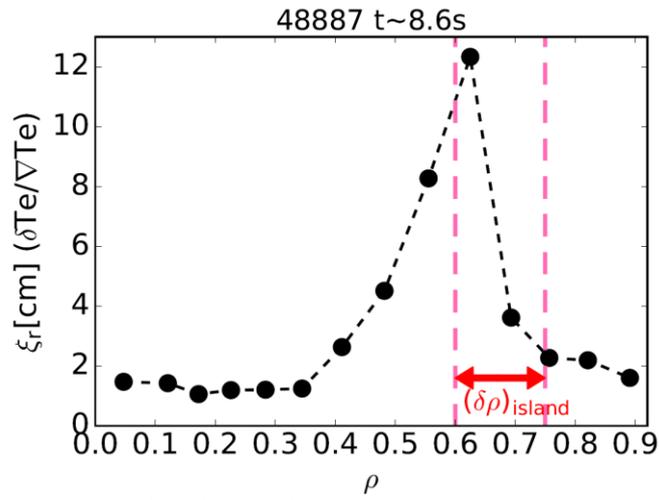

Figure. 8. The profile of the radial displacement ξr estimated with ECE measurement around 8.6 s in 48887 when an $m/n = 2/1$ TM and an $m/n = 1/1$ HC is observed. ξr is estimated as $\delta T_e/\nabla T_e$. Here, the mode frequency of the $m/n = 2/1$ NTM is about 4 Hz. The MHD equilibrium is calculated at 8.66 s. The region where the flattening of the ion temperature profile obtained with CXRS measurement is shown by the red arrow. The width of the $m/n = 2/1$ magnetic island is about $(\delta\rho)_{island} \sim 0.15$.

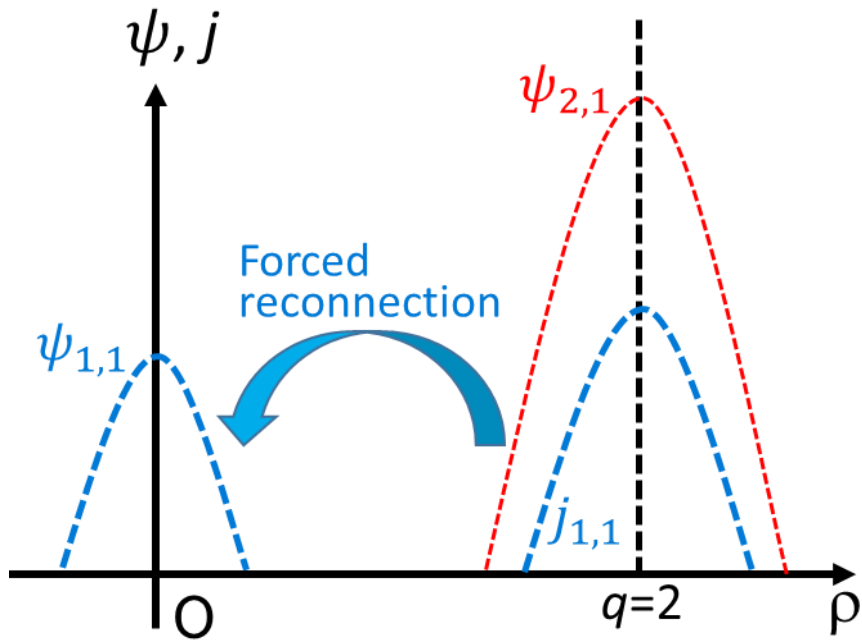

Figure. 9. Schematic view of the excitation mechanism of the $n = 1$ HC by the $m/n = 2/1$ TM. $\psi_{2,1}$ and $\psi_{1,1}$ are $m/n = 2/1$ and $m/n = 1/1$ component of the poloidal flux, respectively. $j_{1,1}$ is $m/n = 1/1$ component of the plasma current as the sideband.

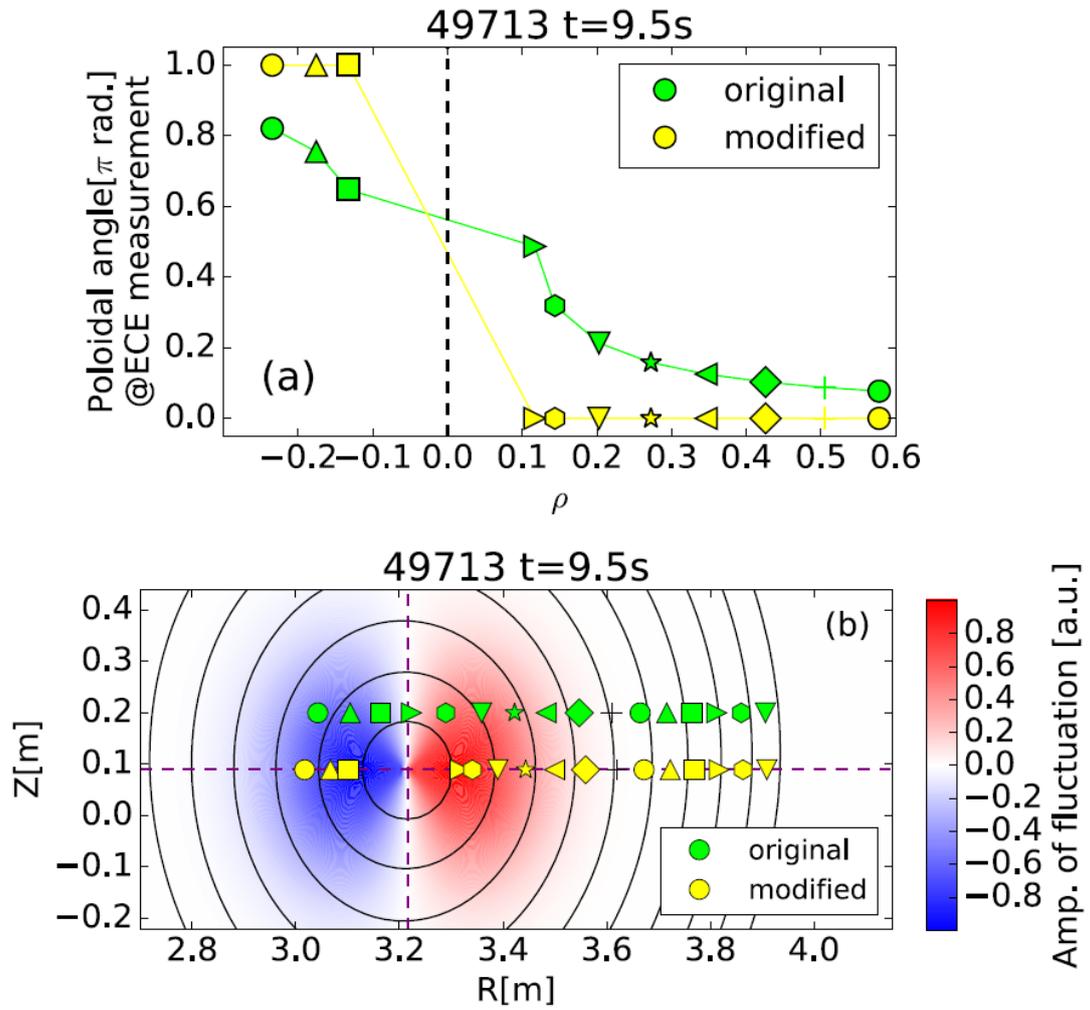

Figure. 10. (a) The radial profile of the poloidal angle measured by ECE measurement. (b) The flux surfaces in a poloidal cross section and the measured positions of ECE measurement. In (b), the contour map shows the amplitude of the fluctuation of an $m = 1$ mode in the core region. In figure (a) and (b), the green and yellow markers mean the original measured poloidal angles (points) by ECE measurement and the modified measured poloidal angles (points) with the method explained in Appendix A, respectively.